\documentclass[preprint]{elsarticle}
\usepackage{amssymb,setstack}

\def\ba{\begin{array}}
\def\ea{\end{array}}
\def\be{\begin{equation}}
\def\ee{\end{equation}}
\def\nn{\nonumber}
\newtheorem{veta}{Theorem}

\def\pd{\partial}

\def\C{{\mathbb C}}
\def\R{{\mathbb R}}

\def\F{{\rm F}}
\def\a{{\mathfrak a}}

\def\g{{\mathfrak g}}

\def\n{{\mathfrak n}}

\def\s{{\mathfrak s}}

\def\z{{\mathfrak z}}
\def\n{{\mathfrak n}}
\def\der{{\mathfrak{Der}}}
\def\inn{{\mathfrak{Inn}}}
\def\na{{\mathfrak n}_{n,3}}
\def\span{{\rm span}}
\def\DS{{\rm DS}}
\def\CS{{\rm CS}}
\def\US{{\rm US}}

\def\ad{{\rm ad}}

\newcommand{\JPA}{{\it J. Phys. A} }
\newcommand{\JMP}{{\it J. Math. Phys.} }
\newcommand{\PL}{{\it Phys. Lett.} }

\begin{document}

\title{Can solvable extensions of a nilpotent subalgebra be useful in the classification of solvable algebras with the given nilradical?}

\author[fjfi]{L. \v Snobl\corref{cor1}}
\ead{Libor.Snobl@fjfi.cvut.cz}
\author[fjfi]{D. Kar\'asek}
\ead{karasda@seznam.cz}

\address[fjfi]{Faculty of Nuclear Sciences and Physical Engineering, \\
Czech Technical University in Prague, B\v rehov\'a 7, 115 19 Prague 1, Czech Republic}

\cortext[cor1]{Corresponding author}

\begin{abstract}
We construct all solvable Lie algebras with a specific $n$--dimensional nilradical $\na$ which contains the previously studied filiform nilpotent algebra $\n_{n-2,1}$ as a subalgebra but not as an ideal. Rather surprisingly it turns out that the classification of such solvable algebras can be reduced to the classification of solvable algebras with the nilradical $\n_{n-2,1}$ together with one additional case. Also the sets of invariants of coadjoint representation of $\na$ and its solvable extensions are deduced from this reduction. In several cases they have polynomial bases, i.e. the invariants of the respective solvable algebra can be chosen to be Casimir invariants in its enveloping algebra.
\end{abstract}


\maketitle

\section{Introduction}\label{intro}

The current article belongs to a series of papers initiated by Pavel Winternitz in \cite{RW} and continued throughout the years with his various collaborators in \cite{NW,NW1,TW,TW1,SW,SW1}. All these papers dealt with the problem of classification of all solvable Lie algebras with the given $n$--dimensional nilradical, e.g. Abelian, Heisenberg algebra, the algebra of strictly upper triangular matrices etc., for arbitrary finite dimension $n$. Other similar series have been recently investigated by different groups in \cite{ACSV} (naturally graded nilradicals with maximal nilindex and a Heisenberg subalgebra of codimension one) and \cite{WLD,WCD} (certain filiform and quasi--filiform nilradicals).

As is well known, the problem of the classification of all solvable (including nilpotent) Lie algebras in an arbitrarily large finite dimension is presently unsolved and is generally believed to be unsolvable. All known full classifications terminate at relatively low dimensions, e.g. the classification of nilpotent algebras is available at most in dimension 8 \cite{Saf,Tsa1}, for the solvable ones in dimension 6 \cite{Mub3,Tur2}.
The unifying idea behind the series \cite{RW,NW,NW1,TW,TW1,SW,SW1} is the belief that the knowledge of full classification of all solvable extensions of certain series of nilradicals can be very useful for both theoretical considerations -- e.g. testing various hypotheses about general structure of solvable Lie algebras -- and practical purposes -- e.g. when a generalization of a given algebra or its nilradical to higher dimensions appears in some physical theory.

In this paper we shall consider the nilradical 
$$ \na=\span \{ x_1,\ldots,x_n \}, \qquad n \geq 5 $$
with the following nonvanishing Lie brackets
\begin{eqnarray}
\nn [x_2,x_n] & = & x_1,  \\
\nn [ x_{3}, x_{n-1} ] & = & x_{1},\\
\label{nla} 
[x_k,x_{n-1}] & = & x_{k-1}, \qquad 4\leq k \leq n-2,\\
\nn [x_{n-1},x_{n}] & = & x_2.
\end{eqnarray}
When $n=5$, the only remaining nonvanishing Lie brackets are 
\begin{equation}\label{nla5}
[x_2,x_n] = [ x_{3}, x_{n-1} ] =  x_1, \quad [x_{n-1},x_{n}]  =  x_2. 
\end{equation}
This $n$--dimensional nilpotent Lie algebra $\na$ is nilpotent of degree of nilpotency (or nilindex, i.e. the highest value of $k$ for which we have $\g^{k} \neq 0$) equal to $n-3$ and with $(n-2)$--dimensional maximal Abelian ideal. It has one--dimensional center $C(\na)=\span\{ x_1 \}$.

Later it will become important for our investigation that it contains as a subalgebra the nilpotent algebra $\n_{n-2,1}$
\begin{eqnarray}
\label{nlaSW1}  [ y_k,y_{n-2}] & = & y_{k-1}, \; 2\leq k \leq n-3,
\end{eqnarray}
 whose solvable extensions were investigated in \cite{SW}. Namely, we have $\tilde{\n}_{n-2,1}$ spanned by $x_1,x_3,\ldots,x_{n-1}$. Similarly, 
$\na$ also contains  $\tilde{\n}_{6,3}$ spanned by $x_1,x_2,x_3,x_4,x_{n-1},x_{n}$. Here, tildes were used to denote these particular embeddings of algebras of the type (\ref{nlaSW1}) and (\ref{nla}), respectively, into the $n$--dimensional nilradical $\na$. We stress that neither $\tilde{\n}_{n-2,1}$ nor $\tilde{\n}_{6,3}$ are ideals. Although in general the knowledge of solvable extensions of a subalgebra of the given nilradical is not of much usefulness in the classification of all solvable extensions of the nilradical for an obvious reason -- the outer derivations of the nilradical need not to leave the subalgebra invariant because it is not invariant even with respect to inner derivations -- we shall see that in the particular case of the nilradical $\na$ considered here all the classification can be brought to the cases of $\n_{n-2,1}$ already investigated in \cite{SW} and $\n_{6,3}$.
\medskip

In the following we shall firstly find out the general form of an automorphism and a derivation of $\na$. Next, we use this knowledge in the construction of all solvable extensions of the nilradical $\na$. Finally, we deduce generalized Casimir invariants of both $\na$ and its solvable extensions.

Throughout the paper we shall use the same notation as in \cite{SW1}. We have attempted to make the present paper self--contained but if any doubts arise about chosen conventions etc. the reader may consult \cite{SW1} as a suitable reference. Also, if the reader desires to get a more general background information about the classification of solvable Lie algebras, the construction of Casimir invariants and so on, we refer him to the review parts of \cite{SW1} and the literature cited there.

\section{Automorphisms and derivations of the nilradical $\na$}

In the computations below we shall assume that $n\geq 7$. The results for $n=5,6$ are derived at the end of Section \ref{classif}.

The nilpotent algebra $\n=\na$ has the following complete flag of ideals
\begin{equation}\label{flag}
 0 \subset \n^{n-3} \subset \n^{n-4} \subset \z_2\subset \z_3\cap \n^2\subset \ldots \subset \z_{n-5} \cap \n^2 \subset \n^2 \subset (\z_2)_{\n} \subset (\n^{n-4})_{\n}\subset \n 
\end{equation} 
where 
\begin{itemize}
\item $\n^{k}$ are elements of the lower central series, defined recursively
\begin{equation}
\n^{1}=\n, \qquad \n^{k} = [\n^{k-1},\n], \ k\geq 2,
\end{equation}
\item $\z_k$ are elements of the upper central series -- that means that $\z_k$ is the unique ideal in $\n$ such that $\z_k/\z_{k-1}$ is the center of $\n/\z_{k-1}$; the recursion starts from the center of $\n$, i.e.  $\z_1 = C(\n)$,
\item and $(\n^{n-4})_{\n}$ is the centralizer of $\n^{n-4}$ in $\n$, i.e. $$(\n^{n-4})_{\n}=\{ x\in \n | [x,y]=0, \; \forall y \in \n^{n-4}\}.$$
\end{itemize}
By construction, the flag (\ref{flag}) is invariant with respect to any automorphism of the Lie algebra $\n$, i.e. in the basis respecting the flag any automorphism will be represented by an upper triangular matrix. Because derivations of $\n$ can be viewed as infinitesimal automorphisms (i.e. elements of the Lie algebra of the matrix Lie group of automorphisms of $\n$), the same triangular form holds also for them.

Therefore, we find it convenient to change the basis $(x_k)$ of $\n$ defined in Eq.~(\ref{nla}) to a seemingly less natural (i.e. Lie brackets appear more cumbersome) basis $(e_k)$ whose essential advantage over the original one is that it respects the flag (\ref{flag}), i.e. 
the $k$--th subspace in the flag is $\span\{e_1,\ldots,e_k\}$ for all $k$. Namely, we take
\begin{equation}\label{basis}
 e_1=x_1, \, e_2= x_3, \, e_3 = x_2, \,  e_4=x_4, \, \ldots \, , e_{n-2}=x_{n-2}, \, e_{n-1}=x_{n}, \, e_{n}=x_{n-1}.
\end{equation}
The nonvanishing Lie brackets now become
\begin{eqnarray}
\nn [e_2,e_{n}] & = & e_1, \\
\nn [ e_{3}, e_{n-1} ] & = & e_{1},\\
\label{nla1} 
[e_4,e_{n}] & = & e_2, \\
\nn [e_k,e_{n}] & = & e_{k-1}, \qquad 5 \leq k \leq n-2,\\
\nn [e_{n-1},e_{n}] & = & -e_3.
\end{eqnarray}
The important subalgebras isomorphic to $\n_{n-2,1},\n_{6,3}$ are now expressed as
$${\tilde\n}_{n-2,1}=\span\{e_1,e_2,e_4,\ldots,e_{n-2},e_{n}\},\qquad {\tilde\n}_{6,3}=\span\{ e_1,e_2,e_3,e_4,e_{n-1},e_{n} \} ,$$
respectively. The ideals in the derived, lower central and upper central series are 
\begin{eqnarray*}
 & & \n^2  = \n^{(1)} = \span \{ e_1,\ldots,e_{n-3} \}, \quad \n^{(2)}=0, \\
 & & \n^k  =  \span \{ e_1,e_2,e_4,\ldots,e_{n-k-1} \}, \,  3\leq k \leq n-5, \\
 & & \n^{n-4} =  \span \{ e_1,e_2 \}, \quad \n^{n-3} = \span \{ e_1 \}, \quad \n^{n-2} = 0,\\
 & & \z_1  =  \n^{n-3}, \quad \z_2  =  \span\{e_1,e_2, e_3\},\\ 
& & \z_k  =  \span\{e_1,\ldots, e_{k+1}, e_{n-1} \}, \, 3\leq k \leq n-4, \quad \z_{n-3}=\n. 
\end{eqnarray*}
 
In order to find the structure of an arbitrary automorphism of $\na$ we consider its matrix in the basis (\ref{basis})
\begin{equation}
 \Phi(e_k)= e_j \Phi_{jk}
\end{equation}
(summation over repeated indices applies throughout the paper unless indicated otherwise). As mentioned above, such matrix must be necessarily upper triangular because the flag (\ref{flag}) is preserved. It is also obvious that the knowledge of its last three columns, i.e. of 
$\Phi(e_{n-2}),\Phi(e_{n-1})$ and $\Phi(e_{n})$, is sufficient for the knowledge of the whole matrix $\Phi$ due to the definition of an automorphism
$$ \Phi([x,y])=[\Phi(x),\Phi(y)], \qquad \forall x,y \in \n$$
and the Lie brackets (\ref{nla1}) -- we can recover all $\Phi(e_k),\, 1\leq k\leq n-3$ through multiple brackets of $\Phi(e_{n-2}),\Phi(e_{n-1})$ and $\Phi(e_{n})$. A natural question is under which conditions do the relations
$$ \Phi(e_{n-2}) = \alpha e_{n-2} + \sum_{k=1}^{n-3} \phi_{k} e_k, $$
$$ \Phi(e_{n-1}) = \beta e_{n-1} + \gamma e_{n-2}+\sum_{k=1}^{n-3} \psi_{k} e_k, $$
$$ \Phi(e_{n}) = \kappa e_{n}  +\lambda e_{n-1} + \mu e_{n-2}+\sum_{k=1}^{n-3} \rho_{k} e_k $$
give rise to an automorphism? 

Obviously, we must have $ \alpha\beta\kappa\neq 0$ to have an invertible map. The preservation of $\z_3$ implies $\gamma=0, \,  \psi_{k}=0,k=5,\ldots,n-3$. The remaining conditions are found as follows
\begin{itemize}
 \item $0=\Phi([e_{n-2},e_{n-1}]) %=\phi_3\beta e_1
$ implies $\phi_3=0$,
 \item $0=\Phi([[e_{n-1},e_{n}],e_n])$ leads to $\psi_4= \frac{\lambda}{\kappa}\beta$,
 \item $ \Phi([[e_{n-1},e_{n}],e_{n-1}])+\Phi\left( (-\\ad_{e_n})^{n-4} e_{n-2} \right)=0$ leads to $\alpha=\beta^2 \kappa^{5-n}$.
\end{itemize}
All other Lie brackets are either used to define $\Phi(e_k),\, 1\leq k\leq n-3$ or are preserved trivially. Therefore, we conclude that any automorphism of $\na$ is uniquely defined in terms of $2n$ parameters which we have denoted $\beta, \kappa, \lambda,$ $\psi_1, \psi_2, \psi_3,$ $\phi_1, \phi_2, \phi_4,\ldots,\phi_{n-3},$ $\rho_{1},\ldots,\rho_{n-3}$ and acts on the generators of the Lie algebra $\na$ in the following way
\begin{eqnarray}
\nn \Phi(e_{n-2}) & = & \beta^2 \kappa^{5-n} e_{n-2} + \sum_{k=4}^{n-3} \phi_{k} e_k+ \phi_{2} e_2+\phi_{1} e_1, \\
\label{genaut}
\Phi(e_{n-1}) & = & \beta e_{n-1} + \frac{\lambda}{\kappa}\beta e_4+\sum_{k=1}^{3} \psi_{k} e_k, \\
\nn \Phi(e_{n}) & = & \kappa e_{n} +\lambda e_{n-1} + \mu e_{n-2}+\sum_{k=1}^{n-3} \rho_{k} e_k. 
\end{eqnarray}
Taking automorphisms infinitesimally close to the unity, i.e. constructing the Lie algebra of the group of automorphisms, we find the most general derivation in the form
\begin{eqnarray}
\nn D(e_{n-2}) & = & (2 c_{n-1}+(5-n) d_n) e_{n-2} + \sum_{k=4}^{n-3} b_{k} e_k+ b_{2} e_2+b_{1} e_1, \\
\label{gender}
D(e_{n-1}) & = & c_{n-1} e_{n-1} + d_{n-1} e_4+\sum_{k=1}^{3} c_{k} e_k, \\
\nn D(e_{n}) & = & \sum_{k=1}^{n} d_{k} e_k. 
\end{eqnarray}
The action of $D$ on the remaining basis elements $e_1,\ldots,e_{n-3}$ is again found using multiple brackets and the defining property of a derivation
$$ D([x,y])=[D(x),y]+[x,D(y)].$$
In the $2n$--dimensional algebra of derivations $\der(\na)$ we have $(n-1)$--dimensional ideal of inner derivations $\inn(\na)$ of the form
\begin{eqnarray}
\nn D(e_{n-2}) & = &  -c_{3} e_{n-3}, \\
\label{innder}
D(e_{n-1}) & = & c_{3} e_3+ c_{1} e_1, \\
\nn D(e_{n}) & = & \sum_{k=1}^{n-3} d_{k} e_k. 
\end{eqnarray}

\section{Construction of solvable Lie algebras with the nilradical $\na$}\label{classif}

Firstly, we recall how the knowledge of automorphisms and derivations of a given nilpotent Lie algebra $\n$ can be employed in the construction  of all solvable Lie algebras $\s$ with the nilradical $\n$. 

Let us consider a basis of $\s$ in the form $(e_1,\ldots,e_n,f_1,\ldots,f_p)$ where $(e_1,\ldots,e_n)$ is a basis of $\n$ with prescribed Lie brackets.  Since $\n$ is an ideal in $\s$ and the derived algebra of $\s$ falls into $\n$ we necessarily have Lie brackets of the form
\begin{equation}\label{solbas}
 [f_a,e_j]  =  (A_a)^k_{j} e_k, \qquad [f_a,f_b]  =  \gamma^j_{ab} e_j. 
\end{equation} 
Furthermore, $\n$ must be a maximal nilpotent ideal, i.e. any nonvanishing linear combination of the matrices $A_a$  must be non--nilpotent.

The algebra $\s$ doesn't change if we transform its basis. Since the structure of $\n$ is fixed we allow only such transformations  that the Lie brackets in $\n$ are not altered, i.e. 
\begin{equation}\label{chanbas}
 e_k \rightarrow \tilde{e}_k = e_j \Phi_{jk}, \qquad f_a\rightarrow \tilde{f}_a = f_{b} \Xi_{ba} + e_k \Psi_{ka}
\end{equation}
where $\Phi$ is a matrix of an automorphism of $\n$ in the original basis $(e_1,\ldots,e_n)$, $\Xi$ is a regular matrix and $\Psi$ is arbitrary.

We represent all non--nilpotent elements $f_a$ in the basis of $\s$ by the corresponding operators in $\der(\n)\subset {\mathfrak{gl}}(\n)$,
\begin{equation}\label{maptoder}
f_a\in\s \rightarrow D_a=\ad_{f_a} |_{\n} \in \der(\n).
\end{equation}
We note that under this mapping of $f_a$'s to outer derivations we loose some information -- from the knowledge of $D_a,D_b$ we can reconstruct the Lie bracket $[f_a,f_b]$ only modulo the kernel of this map, i.e. modulo elements in the center of $\n$. Nevertheless, the construction of all non--equivalent sets of $(D_1,\ldots,D_p)$ is crucial in the construction of all solvable Lie algebras $\s$ with the nilradical $\n$.

Because Eq. (\ref{maptoder}) defines a homomorphism of $\s$ into $\der(\n)$ we can translate the properties of $f_a$'s to $D_a$'s. 
In particular, the commutator of any $D_a,D_b$ must be an inner derivation and no nontrivial linear combination of $D_a$'s can be nilpotent.
That means that  $(D_1,\ldots,D_p)$ must span an Abelian subalgebra $\a$ in the factor algebra $\der(\n)/\inn(\n)$ such that no nonvanishing element of $\a$ is nilpotent. The subalgebras conjugated under an automorphism of $\n$ are equivalent. Therefore, in an abstract formulation we can say that the Lie brackets of solvable extensions of $\n$ are determined modulo elements in the center of $\n$ by conjugacy classes of Abelian subalgebras $\a$ of the factor algebra $\der(\n)/\inn(\n)$ such that no element of $\a$ is represented by a nilpotent operator on $\n$. Now the practical issue is how one can conveniently construct these classes for particular $\n=\na$?

Let us start by considering one additional basis element $f_1\equiv f$, i.e. one derivation $D$.
The elements of $\der(\na)/\inn(\na)$ can be uniquely represented by outer derivations of the form 
\begin{eqnarray}
\nn D(e_{n-2}) & = & (2 c_{n-1}+(5-n) d_n) e_{n-2} + \sum_{k=4}^{n-4} b_{k} e_k+ b_{2} e_2+b_{1} e_1, \\
\label{gendermi}
D(e_{n-1}) & = & c_{n-1} e_{n-1} + d_{n-1} e_4 + c_{3} e_3 + c_{2} e_2, \\
\nn D(e_{n}) & = & d_{n} e_n+d_{n-1} e_{n-1}+d_{n-2} e_{n-2}
\end{eqnarray}
(the action on $e_1,\ldots,e_{n-3}$ follows from the properties of $D$).
Above, a suitable inner derivation (\ref{innder}) was added to an arbitrary derivation, eliminating $n-1$ parameters. 
We mention that the form (\ref{gendermi}) of the representative of the coset $[D]$ is not invariant under conjugation by an automorphism
$$ D\rightarrow D_\Phi=\Phi^{-1} \circ D \circ \Phi $$
so that we may be forced to use a representative $\Phi(D)'$ of the coset $[\Phi(D)]$, different from $\Phi(D)$, after conjugation -- such a change of representative amounts to an addition of an inner derivation and is understood in all simplifications below whenever we employ an automorphism. Due to the triangular shape of $D$ we see that the sought--after Abelian subalgebras are at most two--dimensional since any higher dimensional subalgebra in $\der(\na)/\inn(\na)$ will necessarily involve nonvanishing nilpotent elements.

Next, we find all possible canonical forms of a coset (\ref{gendermi}) up to conjugation by automorphisms and rescaling. In order to reduce the problem to the one already investigated in \cite{SW} we realize that the derivation of the form (\ref{gendermi}) leaves  $${\tilde\n}_{n-2,1}=\span\{e_1,e_2,e_4,\ldots,e_{n-2},e_{n}\}$$
invariant if and only if $d_{n-1}=0$. We conjugate a given derivation $D$ by the automorphism defined by
$$ \Phi(e_{n-2})= e_{n-2}, \quad \Phi(e_{n-1})= e_{n-1}+\frac{d_{n-1}}{d_n-c_{n-1}} e_4, \quad \Phi(e_{n})= e_{n}+\frac{d_{n-1}}{d_n-c_{n-1}} e_{n-1} $$ 
whenever possible, i.e. when $d_n\neq c_{n-1}$. Now we have $\hat{d}_{n-1}=0$, i.e. $D_\Phi\equiv \hat{D}$ leaves $\tilde{\n}_{n-2,1}$ invariant. The case when none of the conjugate derivations $D_\Phi$ leaves $\tilde{\n}_{n-2,1}$ invariant, which necessarily means $d_n= c_{n-1}, \, d_{n-1}\neq 0$, will be dealt with later on, on page \pageref{noninv}.

Provided we set $d_{n-1}= 0$, the outer derivation (\ref{gendermi}) restricted to ${\tilde\n}_{n-2,1}$ has the same structure as in \cite{SW}, Eq. (25). Consequently, we may consider all solvable extensions of ${\tilde\n}_{n-2,1}$ and then extend these to solvable extensions of $\na$, i.e. determine the parameters $c_{n-1},c_{3},c_{2}$. In this way we obtain all solvable extensions of $\na$ except the case $d_n= c_{n-1}, \, d_{n-1}\neq 0$.

The value of the parameter $c_{n-1}$ is fixed by the structure of the solvable extension of ${\tilde\n}_{n-2,1}$, namely in relation to parameters $\alpha,\beta$ introduced below in Theorem \ref{SWth} we have
$$ c_{n-1}=\frac{1}{2} \left( \beta+(n-5)\alpha \right), \qquad d_n = \alpha.$$

When $c_{n-1}\neq 0$ any derivation $D$ can be brought to $D_\phi$ with $c_{2}=0$ using the automorphism $\Phi$ specified by
$$ \Phi(e_{n-2})=e_{n-2}, \quad \Phi(e_{n-1})=e_{n-1}-\frac{c_2}{c_{n-1}} e_{2}, \quad \Phi(e_{n})=e_{n}.$$ 
 When $c_{n-1}=0$ we cannot eliminate nonvanishing $c_{2}$ by any automorphism but we can bring it to $1$ by rescaling of $e_{k}$'s 
provided such scaling  remains available by the structure of the solvable extension of the subalgebra ${\tilde\n}_{n-2,1}$. It turns out that for $c_{n-1}=0$ two non--conjugate extensions of a derivation of ${\tilde\n}_{n-2,1}$ exist, namely those determined by $c_{2}=0,1$.

A similar consideration can be applied also to the parameter $c_{3}$. When $d_{n}\neq 0$ any derivation $D$ can be brought to $D_\phi$ with $c_{3}=0$ using the automorphism $\Phi$ specified by
$$ \Phi(e_{n-2})=e_{n-2}, \quad \Phi(e_{n-1})=e_{n-1}-\frac{c_3}{d_n} e_{3}, \quad \Phi(e_{n})=e_{n}.$$ 
 When $d_{n}=0$ we cannot eliminate nonvanishing $c_{3}$ by any automorphism. Whether or not $c_{3}$ can be rescaled depends on the residual automorphisms still available -- if the diagonal part of automorphisms is completely fixed by the structure of the solvable extension of the subalgebra ${\tilde\n}_{n-2,1}$ nothing can be done, otherwise we can scale $c_{3}$ to $1$ using the automorphism
$$ \Phi(e_{n-2})=e_{n-2}, \quad \Phi(e_{n-1})=e_{n-1}, \quad \Phi(e_{n})=\frac{1}{c_{3}} e_{n}.$$ 

To sum up, the extension to a derivation of the nilradical $\na$ is unique up to a conjugation when $d_{n}\neq 0$ and $c_{n-1}\neq 0$; otherwise, several non--equivalent extensions do exist.

We recall the results of \cite{SW}

\begin{veta}\label{SWth}
Any solvable Lie algebra $\tilde{\s}$ with the nilradical $\n_{m,1}$ has dimension $\dim \tilde{\s}=m+1$, or $\dim \tilde{\s}=m+2$.
Three types of solvable Lie algebras of dimension $\dim \tilde{\s}=m+1$ exist for any $m \geq 4$. They are represented by the following:
\begin{enumerate}
\item $[\tilde{f},\tilde{e}_k]= \left( (m-k-1)\alpha+\beta \right) \tilde{e}_k, \ k \leq m-1, [\tilde{f},\tilde{e}_m]=\alpha \tilde{e}_m.$
The classes of mutually nonisomorphic algebras of this type are
\begin{eqnarray}
\nn \tilde{\s}_{m+1,1}(\beta): & & \alpha=1, \beta \in \F\backslash \{ 0,m-2 \},  \\
\nn DS  & = &  [m+1,m,m-2,0], \ CS=[m+1,m], US=[0], \\
\nn  \tilde\s_{m+1,2}: & & \alpha=1, \beta=0,  \\
\nn  DS  & = &  [m+1,m-1,m-3,0], \ CS=[m+1,m-1], US=[0], \\
\nn  \tilde{\s}_{m+1,3}: & & \alpha=1, \beta=2-m,  \\
\nn  DS  & = &  [m+1,m,m-2,0], \ CS=[m+1,m], US=[1], \\ 
\nn  \tilde{\s}_{m+1,4}: & & \alpha=0, \beta=1,  \\
\nn  DS  & = &  [m+1,m-1,0], \ CS=[m+1,m-1], US=[0]. 
\end{eqnarray}
\item $\tilde{\s}_{m+1,5}: \qquad  [\tilde{f},\tilde{e}_k]=(m-k) \tilde{e}_k, \ k \leq m-1, [\tilde{f},\tilde{e}_m]=\tilde{e}_m+\tilde{e}_{m-1}.$ 
$$ DS  = [m+1,m,m-2,0], \ CS=[m+1,m], \ US=[0].$$
\item $\tilde{\s}_{m+1,6}(a_3,\ldots,a_{m-1}): \, [\tilde{f},\tilde{e}_k]=\tilde{e}_k + \sum_{l=1}^{k-2} a_{k-l+1} \tilde{e}_l, \ k \leq m-1,$  $[f,\tilde{e}_m]=0,$ $a_j \in \F$, at least one $a_j$ satisfies $a_j \neq 0$. \\
Over $\C$: the first nonzero $a_j$ satisfies $a_j=1$.\\
Over $\R$: the first nonzero $a_j$ for even $j$ satisfies $a_j=1$. If all $a_j=0$ for $j$ even, then the first nonzero $a_j$ 
($j$ odd) satisfies $a_j= \pm 1$. We have
\begin{eqnarray}
\nn DS  & = &  [m+1,m-1,0], \ CS=[m+1,m-1], US=[0]. 
\end{eqnarray}
\end{enumerate}
For each $m \geq 4$ precisely one solvable Lie algebra $\tilde{\s}_{m+2}$ of $\dim \tilde{\s}=m+2$ with the nilradical $\n_{m,1}$ exists. It is represented by a basis 
$( \tilde{e}_1, \ldots, \tilde{e}_m,\tilde{f}_1,\tilde{f}_2 )$ and the Lie brackets involving $f_1$ and $f_2$ are
\begin{eqnarray}
\nn [\tilde{f}_1,\tilde{e}_k] & = & (m-1-k)\tilde{e}_k, \ 1 \leq k \leq m-1, \ [\tilde{f}_1,\tilde{e}_m]=\tilde{e}_m, \\
\nn {[\tilde{f}_2,\tilde{e}_k]} & = & \tilde{e}_k, \ 1 \leq k \leq m-1, \ [\tilde{f}_2,\tilde{e}_m]=0, \ [\tilde{f}_1,\tilde{f}_2]=0.
\end{eqnarray}
For this algebra we have
\begin{eqnarray}
\nn DS  & = &  [m+2,m,m-2,0], \ CS=[m+2,m], US=[0] .
\end{eqnarray}
\end{veta}
Above, we used the abbreviations $\DS,\CS$ and $\US$ for (ordered) lists of integers denoting the dimensions of subalgebras in the derived, lower central and upper central series, respectively. We listed the last (then repeated) entry only once (e.g. we write $\CS=[m,m-1]$ rather than $\CS=[m,m-1,m-1,m-1,\ldots]$). 
\medskip

We must point out, however, that there is a caveat in the presented theorem. If we work over the field $\R$ the group of automorphisms of $\n_{n-2,1}$ used in the derivation of Theorem \ref{SWth} in \cite{SW} is slightly larger than the one we have available for the subalgebra $\tilde{\n}_{n-2,1}$, i.e. inherited from automorphisms of $\na$. In other words, the available automorphisms form a group only locally isomorphic to the group of automorphisms of $\n_{n-2,1}$. Namely, the sign of $\alpha=\beta^2 \kappa^{5-n}$ in Eq. (\ref{genaut}) is restricted -- for given $n$ we have ${\rm sgn}\alpha=({\rm sgn}\kappa)^{n-5}$. As a consequence, for our purposes we must for $n$ even consider $[\tilde{f}, \tilde{e}_{m} ]=\tilde{e}_{m} \pm\tilde{e}_{m-1}$ in $\tilde{\s}_{m+1,5}$ ($m=n-2$).
All other results in Theorem \ref{SWth} hold irrespective of this constraint on allowed automorphisms.

The corresponding solvable extensions of the nilradical $\na$ are summarized in Theorem \ref{thm} below.
\medskip

Coming back to the case $d_n= c_{n-1}, d_{n-1}\neq 0$, \label{noninv} we first rescale $D$ to get $d_n= c_{n-1}=1$ and by scaling of $e_{k}$'s we set $d_{n-1}=1$. Using the automorphism 
$$\Phi(e_{n-2})=e_{n-2},\quad \Phi(e_{n-1})=e_{n-1}, \quad \Phi(e_{n})=e_{n}+\frac{d_{n-2} }{n-6} e_{n-2}$$ we get rid of $d_{n-2}$; it is possible since $n \neq 6$. We got $D$ which preserves the subalgebra ${\tilde\n}_{6,3}$. Therefore, it is enough to consider its solvable extensions (with $d_n= c_{n-1}=1$) and then extend these to solvable algebras with the nilradical $\na$. It turns out that such an enlargement is unique up to conjugation, i.e. fully determined by $d_n= c_{n-1}=1, d_{n-1}= 1,  d_{n-2}= 0$, the remaining parameters in Eq. (\ref{gendermi}) vanish.
\medskip

Finally, the two--dimensional Abelian subalgebras $\a$ in $\der(\na)/\inn(\na)$ are easily obtained using the results of the previous analysis. Such subalgebras must contain two linearly independent elements $D'_1,D'_2$, whose diagonal parameters can be chosen to have the values $c_{n-1}=1,d_{n}=-1$ and $c_{n-1}=1,d_{n}=0$, respectively. Due to the chosen values for $D_1$ we can always go over to
$\tilde{D}_1=(D'_1)_\Phi,\tilde{D}_2=(D'_2)_\Phi$ where $\tilde{D}_1$ was diagonalized by a suitable automorphism $\Phi$. The restriction $[\tilde{D}_1,\tilde{D}_2] \in \inn(\na)$ now restricts $\tilde{D}_2$ to be also diagonal. Therefore, all elements of $\a$ act diagonally on $\na$ in the chosen basis and can be expressed e.g. in the basis defined by $D_1\, (c_{n-1}=0,d_{n}=1)$ and $D_2\, (c_{n-1}=1,d_{n}=0)$. The corresponding non--nilpotent elements $f_1,f_2$ in $\s$ in general satisfy
$$ [f_1,f_2]=\alpha e_1 \in C(\n) $$
but a simple redefinition $f_{1}\rightarrow f_1+\frac{\alpha}{2} e_1$ gives an isomorphic solvable algebra $\s$ with $ [f_1,f_2]=0$.\medskip

To sum up, we have the following theorem

\begin{veta}\label{thm}
Any solvable Lie algebra $\s$ with the nilradical $\na$ has dimension $\dim \s=n+1$, or $\dim \s=n+2$.

Five types of solvable Lie algebras of dimension $\dim \s=n+1$ with the nilradical $\na$ exist for any $n \geq 7$. They are represented by the following:
\begin{enumerate}
\item $[ f , e_1] = (\alpha+ 2 \beta)e_1,  \ [ f , e_2]  =  2 \beta e_2, {[f , e_3]} =  (\alpha+ \beta)e_3,$\\
  $ [f , e_k]  =  ((3-k)\alpha + 2 \beta) e_k, \; 4\leq k\leq n-2,$ \\ $[ f, e_{n-1}]  = \beta e_{n-1},  \ [  f, e_n]  =  \alpha e_n.$

The classes of mutually nonisomorphic algebras of this type are
\begin{eqnarray}
\nn \s_{n+1,1}(\beta): & & \alpha=1, \beta \in \F \backslash \{ 0,-\frac{1}{2},\frac{n-5}{2} \},  \\
\nn DS  & = &  [n+1,n,n-3,0], \ CS=[n+1,n], US=[0], \\
\nn \s_{n+1,2}: & & \alpha=1, \beta=\frac{n-5}{2},  \\
\nn  DS  & = & [n+1,n-1,n-4,0], \ CS=[n+1,n-1], US=[0], \\
\nn \s_{n+1,3}: & & \alpha=1, \beta=0,  \\
\nn  DS  & = &  [n+1,n-1,n-4,0], \ CS=[n+1,n-1], US=[0], \\
\nn \s_{n+1,4}: & & \alpha=1, \beta= -\frac{1}{2},  \\
\nn  DS  & = &  [n+1,n,n-3,0], \ CS=[n+1,n], US=[1], \\ 
\nn \s_{n+1,5}: & & \alpha=0, \beta=1,  \\
\nn  DS  & = &  [n+1,n-1,1,0], \ CS=[n+1,n-1], US=[0]. 
\end{eqnarray}
\item $\s_{n+1,6}(\epsilon): \qquad  [f,e_1] = (n-3)e_1,  \ [ f , e_2] = (n-4) e_2,$\\
  $ {[f , e_3]} = (1+ \frac{n-4}{2}) e_3, \, [f , e_k]  =  (n-1-k) e_k, \; 4\leq k\leq n-2,$ \\ $[ f, e_{n-1}]  = \frac{n-4}{2} e_{n-1},  \ [  f, e_n]  =  e_n+\epsilon e_{n-2}$
where $\epsilon=1$ over $\C$, whereas over $\R$ $\epsilon=1$  for $n$ odd, $\epsilon=\pm 1$ for $n$ even.
$$ DS  =  [n+1,n,n-3,0], \ CS=[n+1,n], \ US=[0]. $$ 
\item $\s_{n+1,7}: \qquad [ f , e_1] = e_1,  \ [ f , e_2]  =  0 , {[f , e_3]} =  e_3-e_1,$\\
  $ [f , e_k]  =  (3-k) e_k, \; 4\leq k\leq n-2, \, [ f, e_{n-1}]  = e_2,  \ [  f, e_n]  =  e_n.$
$$ DS  =  [n+1,n-1,n-4,0], \ CS=[n+1,n-1], \ US=[0]. $$ 
\item $\s_{n+1,8}(a_2,a_3,\ldots,a_{n-3}): \qquad  
[f,e_1]= e_1, \, [f,e_2]= e_2,\, [f,e_3]=\frac{1}{2} e_3, $ \\ 
$ [f,e_k]= e_k + \sum_{l=4}^{k-2} a_{k-l+1} e_l+a_{k-2} e_2+ a_{k-1} e_1, \ 4\leq  k \leq n-2,$  \\ 
$ [f,e_{n-1}]= \frac{1}{2} e_{n-1}+a_2 e_3, \, [f,e_n]=0, $\\
$a_j \in \F$, at least one $a_j$ satisfies $a_j \neq 0$. \\
Over $\C$: the first nonzero $a_j$ satisfies $a_j=1$.\\
Over $\R$: the first nonzero $a_j$ for even $j$ satisfies $a_j=1$. If all $a_j=0$ for $j$ even, then the first nonzero $a_j$ 
($j$ odd) satisfies $a_j= \pm 1$. We have
\begin{eqnarray}
\nn DS  & = &  [n+1,n-1,1,0], \ CS=[n+1,n-1], US=[0]. 
\end{eqnarray}
\item $\s_{n+1,9}: \qquad [ f , e_1] = 3e_1,  \ [ f , e_2]  =  2 e_2, {[f , e_3]} =  2e_3-e_2,$\\
  $ [f , e_k]  =  (5-k) e_k, \; 4\leq k\leq n-2,$ \\ $[ f, e_{n-1}]  = e_{n-1}
+e_4,  \ [  f, e_n]  =  e_n+e_{n-1}. $ 
$$ DS  =  [n+1,n,n-3,0], \ CS=[n+1,n], \ US=[0]. $$ 
\end{enumerate}
Precisely one solvable Lie algebra $\s_{n+2}$ of $\dim \s=n+2$ with the nilradical $\na$ exists for any $n \geq 7$. It is presented in a basis 
$( e_1, \ldots, e_n,f_1,f_2 )$ where the Lie brackets involving $f_1$ and $f_2$ are
\begin{eqnarray}
\nn [ a f_1+ b f_2 , e_1] & = & (a+ 2 b)e_1,  \ [ a f_1+ b f_2 , e_2]  =  2 b e_2, \ {[a f_1+ b f_2 , e_3]}  =  (a+ b)e_3, \\
\nn [ a f_1+ b f_2 , e_k] &  = &  ((3-k)a + 2 b)e_k, \; 4\leq k\leq n-2,  \\
\nn {[ a f_1+ b f_2 , e_{n-1}]} & = & b e_{n-1},  \quad [ a f_1+ b f_2 , e_n]  =  a e_n, \quad [f_1,f_2]=0.
\end{eqnarray}
For this algebra we have
\begin{eqnarray}
\nn DS  & = &  [n+2,n,n-3,0], \ CS=[n+2,n], US=[0] .
\end{eqnarray}
\end{veta}
We note that the class $\s_{n+1,8}(a_2,a_3,\ldots,a_{n-3})$ encompasses both extensions of $\tilde{\s}_{m+1,7}(a_3,\ldots,a_{m-1})$ and an
extension of $\tilde{\s}_{m+1,4}$ with $c_{3}\neq 0$ in Eq. (\ref{gendermi}). The choice of the parameter brought to $\pm 1$ was selected in the most convenient form for the presentation and consequently is equivalent but slightly different from the direct extension of $\tilde{\s}_{m+1,7}(a_3,\ldots,a_{m-1})$ to the nilradical $\na$ -- for that choice the non--equivalent values of parameters would be more cumbersome to write down. 
\medskip

When $n=6$ the results are as follows: all the algebras presented in Theorem \ref{thm} exist (with $e_{n-2}\equiv e_{4}$) but they do not exhaust all the possibilities. The reason for this is that in this particular dimension we have $[f , e_{n-2}]= (2 c_5-d_6) e_{n-2}+\ldots $. Therefore, if $d_6=c_5$ then also $[f , e_{n-2}]= d_6 e_{n-2}+\ldots$. That implies that if in the derivation (\ref{gendermi}) we have $d_6=c_5\rightarrow 1, \, d_5\neq 0, \, d_4\neq 0$ then we can set to zero neither $d_5$ nor $d_4$ by any choice of automorphism $\Phi$ and we are left with only one scaling available -- preferably used to set $d_5\rightarrow 1$.

That means that for the $6$--dimensional nilradical $\n_{6,3}$ we have solvable extensions $\s_{7,1}(\beta),\s_{7,2},\s_{7,3},\s_{7,4},\s_{7,5},\s_{7,6}(\epsilon),\s_{7,7},\s_{7,8}(1,a_3),\s_{7,8}(0,\epsilon),\s_{7,9},\s_{8}$ where $\epsilon=1$ over $\C$ and $\epsilon=\pm 1$ over $\R$, whose structure is as described in Theorem \ref{thm} above and one additional class of algebras, differing from $\s_{6,9}$ by one additional nonvanishing parameter $\alpha$ 
\begin{itemize}
 \item $\s_{6,10}(\alpha), \, \alpha \neq 0: \qquad [ f , e_1] = 3e_1,  \ [ f , e_2]  =  2 e_2, {[f , e_3]} =  2e_3-e_2,$\\
  $ [f , e_4]  =  e_4, \quad [ f, e_{5}]  = e_{5} +e_4,  \quad [  f, e_6]  =  e_6+e_{5}+\alpha e_4, $ 
$$ DS  =  [7,6,3,0], \ CS=[7,6], \ US=[0]. $$ 
\end{itemize}
\medskip

When $n=5$, the investigation must be performed in a different way. Namely, there is no $\tilde{\n}_{3,1}$ subalgebra -- it has collapsed
 to the Heisenberg algebra which has different properties. Nevertheless, by a rather straightforward, if repetitive, computation (essentially linear algebra of $5 \times 5$ matrices) one can construct all solvable extensions of $\n_{5,3}$. Since this was done already in \cite{Mub3} for one non--nilpotent element and for two elements the result can be derived from the previous one, we shall only list the results and compare them to their higher dimensional analogues. In order to make the comparison as simple as possible we work in a basis analogous to Eq. (\ref{basis}), namely
\begin{equation}
 e_1=x_1, \, e_2= x_3, \, e_3 = x_2, e_{4}=x_{5}, \, e_{5}=x_{4}.
\end{equation}
The nonvanishing Lie brackets are
\begin{equation}\label{nvlb5}
 [e_2,e_5]=e_1, \; [e_3,e_4]=e_1,\, [e_4,e_5]=-e_3.
\end{equation} 
Although the structure of the nilradical is quite different from the other elements of the series, the set of solvable extensions is rather similar. We get analogues of all solvable algebras in Theorem \ref{thm} with some changes in the structure of $\s_{n+1,6}$, $\s_{n+1,8}$, $\s_{n+1,9}$; in addition, the two algebras $\s_{n+1,2}$ and $\s_{n+1,3}$ become identical when $n=5$. The fact that the algebras $\s_{n+1,6}$, $\s_{n+1,8}$, $\s_{n+1,9}$ must be modified when $n=5$ can be inferred already from Theorem \ref{thm} since the Lie brackets as presented there cannot be made sense of if $n=5$. These structurally different analogues are distinguished by primes below.

Explicitly, assuming the structure of $\n_{5,3}$ in the form (\ref{nvlb5}), we have the following Lie brackets with non--nilpotent element(s) and dimensions of the characteristic series
\begin{itemize}
\item $\s_{6,1}(\beta),$ $\beta \in \F \backslash \{ 0,-\frac{1}{2} \}:$\\
%G6_94
$[f,e_1]=(1+2\beta)e_1, \, [f,e_2]=2\beta e_2, \, [f,e_3]=(\beta+1)e_3, \, [f,e_4]=\beta e_4, \, [f,e_5]= e_5$,
$$DS  =   [6,5,2,0], \; CS=[6,5], \; US=[0].$$
\item  $\s_{6,2}:$  
%G6_980
$[f,e_1]=e_1, \, [f,e_2]=0, \, [f,e_3]=e_3, \, [f,e_4]=0, \, [f,e_5]= e_5$, 
$$DS   =  [6,3,0], \; CS=[6,3], \; US=[0].$$
\item $\s_{6,4}:$ 
%G6_942 rescaled
$[f,e_1]=0, \, [f,e_2]=- e_2, \, [f,e_3]=\frac{1}{2} e_3, \, [f,e_4]= -\frac{1}{2} e_4, \, [f,e_5]= e_5$, 
$$DS   =   [6,5,2,0], \; CS=[6,5],\; US=[1].$$
\item $\s_{6,5}:$ 
%G6_940sigma=0
$[f,e_1]=2 e_1, \, [f,e_2]=2 e_2, \, [f,e_3]= e_3, \, [f,e_4]=e_4, \, [f,e_5]= 0$, \\
$$DS  =  [6,4,1,0], \; CS=[6,4], \; US=[0].$$ 
\item $\s'_{6,6}:$
%g6_97
 $[f,e_1] = 2e_1,  \ [ f , e_2] = e_2, {[f , e_3]} = \frac{3}{2} e_3,\ [ f, e_{4}]  = \frac{1}{2} e_{4},  \ [  f, e_5]  =  e_5+ e_{2},$
$$ DS  =  [6,5,2,0], \ CS=[6,5], \ US=[0]. $$ 
\item $\s_{6,7}:$ 
% g6_98
$[ f , e_1] = e_1,  \ [ f , e_2]  =  0 , {[f , e_3]} =  e_3-e_1, \ [ f, e_4]  = e_2,  \ [  f, e_5]  =  e_5,$
$$ DS  =  [6,4,1,0], \ CS=[6,4,3], \ US=[0]. $$ 
\item $\s'_{6,8}:$
%g6_940sigma=1
$ [f,e_1]= 2 e_1, \, [f,e_2]= 2 e_2, \,  [f,e_3]=e_3, [f,e_{4}]= -e_{3}+e_4, \, [f,e_5]=0, $
$$ DS   =   [6,4,1,0], \ CS=[6,4], US=[0]. $$
\item $\s'_{6,9}:$
% g6_96
$[f,e_1]=3e_1, \, [f,e_2]=2e_2-e_3, \, [f,e_3]=2e_3, \, [f,e_4]= e_4+e_5, \, [f,e_5]= e_5,$  
$$ DS  =  [6,5,2,0], \ CS=[6,5], \ US=[0]. $$ 
\item $\s_{7}:$ 
% G7
$[f_1,e_1]=e_1, \, [f_1,e_2]=0, \, [f_1,e_3]=e_3, \, [f_1,e_4]=0, \, [f_1,e_5]= e_5$, \\
$[f_2,e_1]=2e_1, \, [f_2,e_2]=2e_2, \, [f_2,e_3]=e_3, \, [f_2,e_4]=e_4, \, [f_2,e_5]= e_5$, \\
$$ DS  =  [7,5,2,0], \; CS=[7,5], \; US=[0].$$
\end{itemize}
We note that in several cases the characteristic series are different from the ones in Theorem \ref{thm}. This difference in behavior is due to the structural difference between $\n_{n-2,1}$ and the Heisenberg algebra.

\section{Generalized Casimir invariants}

We proceed to construct the generalized Casimir invariants, i.e. the invariants of the coadjoint representation, of the nilpotent algebra $\na$ and its solvable extensions. We recall that a basis for the coadjoint representation of the Lie algebra $\g$ is given by the first order differential operators
\be\label{doal}
 \hat X_k = x_a c^a_{kb} \frac{\pd}{\pd x_b}
\ee
acting on linear functions on $\g$, i.e. elements of $\g^*$. Here, $c^k_{ij}$ are the structure constants of Lie algebra $\g$ in the basis $(x_1,\ldots,x_N)$. In Eq. (\ref{doal}) the quantities $x_a$ are commuting independent variables which are identified with coordinates in the basis of the space $\g^*$ dual to the basis $(x_1,\ldots,x_N)$ of the algebra $\g$.

The invariants of the coadjoint representation, i.e. the generalized Casimir invariants, are solutions of the following system of partial differential equations
\be\label{casimir}
 \hat X_k I(x_1, \ldots,x_N)=0, \ k=1,\ldots,N .
\ee
Several methods exist for the construction of the invariants of the coadjoint representation, most widely used ones are direct solution of Eq. (\ref{casimir}) by the method of characteristics (see e.g. \cite{Abellanas, Abellanas1,PSW}) and the method of moving frames (see \cite{CartanMF1,CartanMF2,FelsOlver1,FelsOlver2,BPP1,BPP2}). 

However, we shall use a different approach. We reduce the equations (\ref{casimir}) to the ones encountered and solved in \cite{SW} for the subalgebra $\tilde{\n}_{n-2,1}$ and its solvable extensions. 

Considering first the nilpotent algebra $\na$ we have the operators (\ref{doal}) in the form
\begin{eqnarray}
\nn \hat E_1 & = & 0, \ \, \hat E_2  =  e_{1} \frac{\pd}{\pd e_n}, \ \ E_3  =  e_{1} \frac{\pd}{\pd e_{n-1}}, \ \, \hat E_4  =  e_{2} \frac{\pd}{\pd e_n}, \\
\label{doal1} \hat E_k & = & e_{k-1} \frac{\pd}{\pd e_n}, \; 5\leq k\leq n-2, \ \, \hat E_{n-1}  =  -e_{1} \frac{\pd}{\pd e_3}-e_{3} \frac{\pd}{\pd e_n} ,\\ 
\nn \hat E_n & = &  - e_{1} \frac{\pd}{\pd e_2} - e_{2} \frac{\pd}{\pd e_4} -\sum_{k=5}^{n-2} e_{k-1} \frac{\pd}{\pd e_k}+e_3 \frac{\pd}{\pd e_{n-1}}.
\end{eqnarray}
It is evident that any solution $I$ of Eq. (\ref{casimir}) cannot depend\footnote{neither can $I$ depend on $e_{n}$} on $e_3,e_{n-1}$ because of $\hat E_{n-1} I= \hat E_{3} I= \hat E_{2} I=0$. Consequently, all considered operators $\hat E_j$ can be truncated to act on functions of $\tilde{e}_1=e_1,\tilde{e}_2=e_2,\tilde{e}_3=e_4,\ldots,\tilde{e}_{n-3}=e_{n-2},\tilde{e}_{n-2}=e_{n}$ only. Then $\hat E_{3T}, \hat E_{n-1T}$ vanish and the remaining operators are exactly those present in the investigation of invariants of $\n_{n-2,1}$ in \cite{SW}. Therefore, the generalized Casimir invariants of $\na$ are the same as the ones for $\n_{n-2,1}$ once written in an appropriate basis.

Similarly, when we consider the solvable extensions of $\na$, the operators $\hat E_j$ in (\ref{doal1}) get additional $\frac{\pd}{\pd f}$ or $\frac{\pd}{\pd f_1}, \frac{\pd}{\pd f_2}$ terms and one ($\hat F$) or two ($\hat F_1,\hat F_2$) additional operators are present in Eq. (\ref{casimir}). 

Let us first consider the case with $\hat F$ only. When the derivation $D$ defining $f$ is such that $ 2 c_{n-1}+d_n \neq 0$, we have 
$\hat E_1= (2 c_{n-1}+d_n) e_1 \frac{\pd}{\pd f}$ which excludes the dependence of $I$ on $f$. When $2 c_{n-1}+d_n = 0$ the situation is only slightly more complicated -- the operators $\hat E_2,\hat E_4$ together again exclude the dependence of $I$ on both $f$ and $e_n$. In both cases, we can restrict all operators (\ref{doal1}) and $\hat F$ to $\na$ and then to $\n_{n-2,1}$, reducing the computation to the corresponding solvable extension of $\n_{n-2,1}$. 

In the second case we have two additional operators $\hat F_1,\hat F_2$ and $\frac{\pd}{\pd f_1}, \frac{\pd}{\pd f_2}$ terms in $\hat E_j$. Now the operators $\hat E_1, \hat E_2, \hat E_3, \hat E_4$ are used in the same way to show that any invariant $I$ cannot depend on $f_1,f_2$.

Altogether, the construction of generalized Casimir invariants was fully reduced to the one for the nilradical $\n_{n-2,1}$.

We recall the results of \cite{SW}

\begin{veta}\label{SWth2}
The nilpotent Lie algebra $\n_{m,1}$ has $m-2$ functionally independent invariants. They can be chosen to be the following polynomials
\begin{eqnarray}
\nn {\tilde\xi}_0 & = & \tilde{e}_1, \\
 {\tilde\xi}_k & = & \frac{(-1)^k k}{(k+1)!} \tilde{e}_2^{k+1} + \sum_{j=0}^{k-1} (-1)^j \frac{\tilde{e}_2^j \ \tilde{e}_{k+2-j} \ \tilde{e}_1^{k-j}}{j!}, 
\; 1 \leq k \leq m-3.
\end{eqnarray}
The algebras $\tilde{s}_{m+1,1}(\beta),\ldots,\tilde{\s}_{m+1,5}$ have $m-3$ invariants each. Their form is
\begin{enumerate}
\item $\tilde{\s}_{m+1,1}(\beta)$, $\tilde{\s}_{m+1,2}$ and $\tilde{\s}_{m+1,5}$
\be
{\tilde\chi}_k = \frac{{\tilde\xi}_k}{{\tilde\xi}_0^{(k+1)\frac{m-3+\beta}{m-2+\beta}}}, \ 1 \leq k \leq m-3.
\ee
For $\tilde{\s}_{m+1,2}$ and $\tilde{\s}_{m+1,5}$ we have $\beta=0$ and $\beta=1$, respectively in Equation (\ref{sn111215inv}).
\item $\tilde{\s}_{m+1,3}$
\be
{\tilde\chi}_1 = {\tilde\xi}_0, \  {\tilde\chi}_k = \frac{{\tilde\xi}^2_{k}}{{\tilde\xi}_1^{k+1}}, \  2 \leq k \leq m-3.
\ee
\item ${\tilde\s}_{m+1,4}$
\be
{\tilde\chi}_k = \frac{{\tilde\xi}_{k}}{{\tilde\xi}_0^{k+1}}, \  1 \leq k \leq m-3.
\ee
\item $\tilde{\s}_{m+1,7}(a_3,\ldots,a_{m-1})$
\begin{eqnarray}
 {\tilde\chi}_k & = & \sum_{q=0}^{[\frac{k+1}{2}]} (-1)^q \frac{(\ln {\tilde\xi}_0)^q}{q!} \left( \sum_{i_1+\ldots+i_q=k-2q+1} a_{i_1+3}  
a_{i_2+3}\ldots a_{i_q+3} \right. \\
\nn & + & \left. \sum_{j+ i_1+\ldots+i_q=k-2q-1} \frac{{\tilde\xi}_{j+1}}{{\tilde\xi}_0^{j+2}} \ a_{i_1+3}  a_{i_2+3}\ldots a_{i_q+3} \right),
\ 1 \leq k \leq m-3.
\end{eqnarray}
The summation indices take the values $0 \leq j,i_1,\ldots,i_q \leq k+1$.
\end{enumerate}
The Lie algebra $\tilde{\s}_{m+2}$ has $m-4$ functionally independent invariants that can be chosen to be 
\be
{\tilde\chi}_k = \frac{{\tilde\xi}_{k+1}}{{\tilde\xi}_1^{\frac{k+2}{2}}}, \; 1 \leq k \leq m-4  .
\ee
\end{veta}

The results for $\na$ and its solvable extensions are now as follows.
\begin{veta}\label{th2}
Let $n\geq 6$.The nilpotent Lie algebra $\na$ has $n-4$ functionally independent invariants. They can be chosen to be the following polynomials
\begin{eqnarray}
\nn \xi_0 & = & e_1, \\
\label{nac} \xi_k & = & \frac{(-1)^k k}{(k+1)!} e_2^{k+1} + \sum_{j=0}^{k-1} (-1)^j \frac{e_2^j \ e_{k+3-j} \ e_1^{k-j}}{j!}, 
\; 1 \leq k \leq n-5.
\end{eqnarray}
The algebras $\s_{n+1,1}(\beta),\ldots,\s_{n+1,9}$ have $n-5$ invariants each. Their form is
\begin{enumerate}
\item $\s_{n+1,1}(\beta)$, $\s_{n+1,2}$, $\s_{n+1,3}$,  $\s_{n+1,6}$, $\s_{n+1,7}$  and  $\s_{n+1,9}$
\be\label{sn111215inv}
\chi_k = \frac{\xi_k}{\xi_0^{(k+1)\frac{2\beta}{1+2\beta}}}, \ 1 \leq k \leq n-5.
\ee
For $\s_{n+1,2}$ is $\beta=\frac{n-5}{2}$, for $\s_{n+1,3}$  and $\s_{n+1,7}$ we have $\beta=0$, for $\s_{n+1,6}(\epsilon)$ we have $\beta=\frac{n-4}{2}$ and for $\s_{n+1,9}$ is $\beta=1$, respectively in Equation (\ref{sn111215inv}).
\item $\s_{n+1,4}$
\be
\chi_1 = \xi_0, \  \chi_k = \frac{\xi^2_{k}}{\xi_1^{k+1}}, \  2 \leq k \leq n-5.
\ee
\item $\s_{n+1,5}$
\be
\chi_k = \frac{\xi_{k}}{\xi_0^{k+1}}, \  1 \leq k \leq n-5.
\ee
\item $\s_{n+1,8}(a_2,a_3,\ldots,a_{n-3})$
\begin{eqnarray}
 \chi_k & = & \sum_{q=0}^{[\frac{k+1}{2}]} (-1)^q \frac{(\ln \xi_0)^q}{q!} \left( \sum_{i_1+\ldots+i_q=k-2q+1} a_{i_1+3}  
a_{i_2+3}\ldots a_{i_q+3} \right. \\
\nn & + & \left. \sum_{j+ i_1+\ldots+i_q=k-2q-1} \frac{\xi_{j+1}}{\xi_0^{j+2}} \ a_{i_1+3}  a_{i_2+3}\ldots a_{i_q+3} \right),
\ 1 \leq k \leq n-5.
\end{eqnarray}
The summation indices take the values $0 \leq j,i_1,\ldots,i_q \leq k+1$.
\end{enumerate}
When $n=6$ the Lie algebra $\s_{7,10}(\alpha)$ has one invariant which can be chosen in the form $\frac{2 e_4 e_1-e_2^2}{e_1^{4/3}}$, i.e. coincides with the one for $\s_{7,9}$.\medskip

The Lie algebra $\s_{m+2}$ has $n-6$ functionally independent invariants that can be chosen to be 
\be
\chi_k = \frac{\xi_{k+1}}{\xi_1^{\frac{k+2}{2}}}, \; 1 \leq k \leq n-6  .
\ee
\end{veta}
We point out that the algebras $\s_{n+1,3}$  and $\s_{n+1,7}$ are examples of solvable non--nilpotent Lie algebras with polynomial basis of invariants, i.e. their basis of invariants can be chosen in the form of Casimir operators in the enveloping algebra of $\s_{n+1,3}$  and $\s_{n+1,7}$ (the same holds also for $\tilde{s}_{m+1,1}(3-m)$ of \cite{SW}). If ever a hypothesis concerning a criterion for the existence of polynomial basis of invariants of solvable algebras is presented, these examples can be easily used as simple tests of its plausibility.
\medskip

For $5$--dimensional nilradical $\n_{5,3}$ we have solvable algebras $\s_{6,1}(\beta)$, $\s_{6,2}$, $\s_{6,5}$, $\s'_{6,6}$, $\s_{6,7}$, $\s'_{6,8}$, $\s'_{6,9}$ with no invariants and $\s_{6,4}$ which has two invariants. They can be chosen in the polynomial form 
$$e_1, \qquad 2 e_1^2 f -2 e_1 e_2 e_5+e_1 e_3 e_4+e_2 e_3^2.$$
The algebra $\s_{7}$ has one invariant
$$ \frac{(f_2-2f_1) e_1^2+ (2e_2 e_5-e_3 e_4) e_1-e_2 e_3^2}{e_1^2}.$$
We observe that invariants of the solvable Lie algebras with the nilradical $\n_{5,3}$ (if nonconstant) depend on the elements outside of $\n_{5,3}$, i.e. $f$ or $f_1,f_2$. This is related to the fact that there is no $\tilde\n_{3,1}$ subalgebra -- it degenerated to Heisenberg algebra whose properties are markedly different.

\section{Conclusions}

We have fully classified all solvable Lie algebras with the nilradical $\na$ in arbitrary dimension $n$ and constructed their generalized Casimir invariants. 

There are two general lessons to be learned from this computation. Firstly, it turned out that the knowledge of all solvable extensions of a suitable subalgebra $\tilde\n$ of the given nilpotent algebra $\n$ may lead to a significant simplification of the whole computation and is definitively worth investigating if such subalgebras are identified in $\n$. This can hold notwithstanding the fact that not all automorphisms of $\n$ preserve the subalgebra $\tilde\n$. Of course, it was important in our investigation that the structure of the subalgebra was restrictive enough, i.e. we expect that similar simplification can be achieved probably for subalgebras with high enough degree of nilpotency, e.g. filiform or quasi--filiform.

Secondly, it was of profound importance that (almost) all automorphisms of $\tilde\n$ could be obtained as a restriction of automorphisms of $\n$. In our case we had a local isomorphism of $Aut(\tilde\n)$ and $Aut(\n)|_{\tilde\n}$; the two differ topologically by the absence of some connected components of $Aut(\tilde\n)$ in $Aut(\n)|_{\tilde\n}$. This minor difference could be easily taken into account and the classification of all solvable extensions of $\tilde\n$ with respect to this restricted group of automorphisms acting on $\tilde\n$ was obtained by inspection from previously known results \cite{SW}. On the other hand, had the $Aut(\tilde\n)$ and $Aut(\n)|_{\tilde\n}$ been locally non--isomorphic, the knowledge of solvable extensions of $\tilde\n$ would not be of much use in the study of solvable extensions of $\n$. A simple example of this is the maximal Abelian ideal $\a$ of $\n$ -- its group of automorphisms {\it per se} is typically much larger than the automorphisms inherited from $\n$, i.e. many transformations used in $\a$ are not allowed in $\n$ and, at the same time, most of solvable extensions of $\a$ cannot be enlarged to solvable extensions of $\n$ -- the Lie brackets of $\n$ simply don't allow that. Therefore, the particular properties of the subalgebra and its immersion into the whole nilradical are of crucial importance for the whole setup to work.

Finally, we have seen that although the considered series of nilpotent algebras could be rather naturally constructed starting from dimension $n=5$, the 5--dimensional one has substantially different properties which reflect themselves also in the possible solvable extensions and their invariants.
\medskip

{\bf Acknowledgments:} The research of Libor \v Snobl was supported by the research plan MSM6840770039 and the project LC527 of the Ministry of Education of the Czech Republic.  We are grateful to Pavel Winternitz for numerous discussions which led to the results presented in this paper.

\end{document}